\documentclass[twocolumn,english,letterpaper,showpacs]{revtex4}
\usepackage{babel,amsmath,amssymb}
\usepackage[dvips]{graphics}

\begin{document}

\title{Field propagation in the Schwarzschild-de Sitter black hole}

\author{C. Molina}
\email{cmolina@fma.if.usp.br}

\author{E. Abdalla}
\email{eabdalla@fma.if.usp.br}
\affiliation{Instituto de F\'{\i}sica, Universidade de S\~{a}o Paulo\\
C.P. 66318, 05315-970 S\~{a}o Paulo-SP, Brazil}

\author{A. Saa}
\email{asaa@ime.unicamp.br}
\affiliation{Departamento de Matem\'atica Aplicada, UNICAMP \\
C.P.6065, 13083-859 Campinas-SP, Brazil}


\begin{abstract}
We present an exhaustive analysis of scalar, electromagnetic and
gravitational perturbations in the background of a Schwarzchild-de
Sitter spacetime. The field propagation is considered by means of a
semi-analytical (WKB) approach and two numerical schemes: the
characteristic and general initial value integrations. The results are
compared  near the extreme regime, and a unifying picture is
established  for the dynamics of different spin fields.
Although some of the results just confirm usual expectations,
a few surprises turn out to appear, as the dependence on the
non-characteristic initial conditions of the non-vanishing asymptotic
value for $\ell=0$ mode scalar fields. 
\end{abstract}

\maketitle

\section{Introduction}

Wave  propagation around non trivial solutions of Einstein equations, 
black hole in particular, is an active field of research (see
\cite{Regge-57,Chandrasekhar,Kokkotas-99} and references therein). 
The perspective of gravitational waves detection in a near future and
the great development of numerical general relativity 
have increased even more the activity on this field. Gravitational
waves should be especially strong when emitted by black holes. The
study of the propagation of perturbations around them is, hence,
essential to provide templates for the gravitational waves identification.
In the other hand, recent astrophysical
observations indicate that the universe is undergoing an accelerated 
expansion phase, suggesting the existence of a small positive  
cosmological constant and that de Sitter (dS) geometry provides 
a good description of very large scales of the universe
\cite{accel-exp}. We notice also that string theory
has recently motivated many works on asymptotically anti-de Sitter
spacetimes (see, for instance,
\cite{Cardoso-01,AdS-4d,AdS-other,Wang-01}).      

In this work, we perform an exhaustive investigation
of scalar, electromagnetic and gravitational
perturbations in the background of a Schwarzchild-de Sitter spacetime.
We scan the full range of the
cosmological constant, from the asymptotic flat case
($\Lambda=0$) up to the critical value of $\Lambda$
which characterizes the Nariai solution \cite{Nariai}.
Two different numerical methods and a higher order WKB analysis are
used.

We remind that for any perturbation in the spacetimes we consider,
after the initial transient phase there are two main
contributions to the resulting asymptotic wave \cite{Leaver-86,Ching-95}: 
initially the so called quasinormal modes, which are suppressed at later
time by the tails.
The first can be understood as candidates to
normal modes which, however, decay (their energy eigenvalues becomes
complex), as in the ingenious mechanism first described by G. Gamow in
the context of nuclear physics \cite{Gamow-28}. After the initial
transient phase, the properties of resulting wave
are more related to background spacetime rather than to the source itself. 

It is well known that for
 asymptotic flat backgrounds the tails decay
according to a power-law, whereas in a space with a positive cosmological
constant the decay is exponential. Curiously, 
$\ell=0$ modes for scalar fields in dS spacetimes, contrasting
to the asymptotic flat cases, approach exponentially 
a non vanishing asymptotic value \cite{Brady-97,Brady-99}. 
We detected, by using a non-characteristic
numerical integration scheme, a dependence of this asymptotic value
on the initial velocities. In particular, it vanishes for static
initial conditions.

The semi-analytical analyses of this work were performed by using 
the higher order WKB method proposed by Schutz and Will
\cite{Schutz-85}, and improved by Iyer and Will \cite{Iyer-87,WKB-ap}.
It provides a very
accurate and systematic way to study black hole quasinormal modes. 
We apply it to the study of various perturbation fields in the
non-asymptotically flat dS geometry. Quasinormal modes are 
also calculated according to this approximation, and  the results 
are compared to the numerical ones whenever
appropriate, providing a quite complete picture of the question of
quasinormal perturbations for dS black holes.

Two very recent works overlap our analysis presented here.
A similar WKB analysis \cite{Konoplya-03} was done very recently by
Zhidenko \cite{Zhidenko}, and his results coincide with ours.
Yoshida and Futamase \cite{Yoshida} used a continued fraction
numerical code to calculate quasinormal mode frequencies, with special
emphasis to high order modes. Our results are also compatible.
Finally, we notice that solutions of the wave equation in a
non trivial background has also been used to infer intrinsic
properties of the spacetime \cite{qnmbh}. 

\section{Metric, fields and effective potentials}

The metric describing a spherically black hole in the presence of
a cosmological constant is well known in the literature. Written in
spherical coordinates, the Schwarzschild-de Sitter metric is given by
\begin{equation}
ds^{2}=-h(r)dt^{2}+h(r)^{-1}dr^{2}+r^{2}\left( d\theta^2 +
\sin^2\theta d\phi^2  \right) \ ,
\label{metric}
\end{equation}
where the function $h(r)$ is
\begin{equation}
h(r)=1-\frac{2m}{r}-\frac{\Lambda r^{2}}{3} \ .
\end{equation}
The integration constant $m$ is the black hole mass, and if the
cosmological constant is positive, the spacetime is asymptotically de
Sitter. In this case, $\Lambda$ is usually written in terms of a
``cosmological radius'' $a$ as $\Lambda = 3/a^{2}$.

Assuming $m > 0$ and  $0 < 9 m^{2}
\Lambda < 1$, the function
$h(r)$ has two positive zeros $r_{+}$ and $r_{c}$ and a negative zero
$r_{n} = -(r_{+} + r_{c})$. This is the Schwarzschild-de
Sitter (SdS) geometry, in which we are interested. 
The horizons $r_{+}$ and $r_{c}$ are denoted event and
cosmological horizons respectively. In this case, the constants $m$
and $a$ are related with the roots by 
\begin{equation}
a^2 = r_{+}^2 + r_{c}^2 + r_{+} r_{c}\ ,
\end{equation}
\begin{equation}
2 m a^2 = r_{+} r_{c} (r_{+} + r_{c})\ .
\end{equation}
If $9 m^{2} \Lambda = 1$, the zeros $r_{+}$ and $r_{c}$ degenerate in a
double root. This is the extreme SdS black hole. If $9 m^{2} \Lambda > 1$,
there is no real positive zeros, and the metric (\ref{metric}) does
not describe a black hole. 

We will consider scalar, electromagnetic and
gravitational perturbations in the submanifold given by
\begin{equation}
T_{+}=\left\{
(t,r,\theta,\phi),r_{+}<r<r_{c}\right\} \  .
\end{equation}
In this region, we define the ``tortoise'' radial coordinate by
\begin{equation}
x(r) = \frac{1}{2\kappa_{c}} \ln(r_{c} - r) 
       - \frac{1}{2\kappa_{+}} \ln(r - r_{+}) 
       + \frac{1}{2\kappa_{n}} \ln(r - r_{n}) \ ,
\end{equation}
with
\begin{equation}
\kappa_{i} = \frac{1}{2} \left| \frac{d h(r)}{dr}\right|_{r=r_{i}} \ .
\end{equation}
The constants $\kappa_{+}$ and $\kappa_{c}$ are the surface gravity
associated with the event and cosmological horizons.

For a scalar field $\Phi$ obeying the massless Klein-Gordon equation 
\begin{equation}
\Box\Phi = 0 \ , \label{boxphiequalzero}
\end{equation}
the usual separation of variables in terms of a radial field and a
spherical harmonic $\textrm{Y}_{\ell,m}(\theta,\varphi)$,
\begin{equation}
\Phi=\sum_{\ell\,
m} \frac{1}{r} \psi_{\ell}^{sc}(t,r)\textrm{Y}_{\ell}(\theta,\phi)  \ ,
\label{Ansazs_field}
\end{equation}
leads to 
Schr\"{o}dinger-type equations in the tortoise coordinate
for each value of $\ell$,
\begin{equation}
-\frac{\partial^{2}\psi_{\ell}^{sc}}{\partial
 t^{2}}+\frac{\partial^{2}\psi_{\ell}^{sc}}{\partial
 x^{2}}=V^{sc}(x)\psi_{\ell} \ ,
\label{wave_equation}
\end{equation}
where the effective potential $V^{sc}$ is given by
\begin{equation}
V^{sc}(r)=h(r)\left[\frac{\ell(\ell + 1)}{r^2} + \frac{2m}{r^{3}} 
- \frac{2}{a^{2}}\right] \ .
\end{equation}

In the Schwarzschild-de Sitter geometry, in contrast to the the case
of a electrically charged black hole,  it is possible to have pure
electromagnetic and gravitational perturbations. For the first, the
potential of the corresponding Schr\"{o}dinger-type  equation 
is given \cite{Ruffini-72} 
\begin{equation}
V^{el}(r) = h(r)\frac{\ell(\ell + 1)}{r^{2}},
\end{equation}
with $\ell\ge 1$.
The gravitational 
perturbation theory for the exterior  Schwarzschild-de Sitter  geometry
has been developed by \cite{Chandrasekhar,Cardoso-01}. The potential
for the axial and polar modes are, respectively,
\begin{equation}
V^{ax}(r) =
h(r)\left[\frac{\ell(\ell+1)}{r^{2}}-\frac{6m}{r^{3}}\right] \ , 
\end{equation}
\begin{eqnarray}
V^{po}(r) & = & \frac{2 h(r)}{r^{3} \left(3m+cr\right)^{2}} 
\left[ 9m^{3}+3c^{2}mr^{2}+c^{2}(1+c)r^{3} \right. \nonumber \\
& & \mbox{} \left. +3m^{2} (3cr-\Lambda r^{3}) \right]  \ ,
\end{eqnarray}
with $2c = (\ell-1)(\ell+2)$ and $\ell\ge 2$.

For perturbations with $\ell > 0$, we can show explicitly that all the
effective potentials $V(x) \equiv V(r(x))$ are positive definite. For
scalar perturbations with $\ell = 0$, however, the effective potential has one
zero point $x_{0}$  and it is negative for $x > x_{0}$.

\section{Numerical and Semi-analytical approaches}

\subsection{Characteristic integration}

In the work \cite{Gundlach-94} a very simple but at the
same time very efficient way of dealing with two-dimensional
d'Alembertians  has been set up. Along the general lines of the
pioneering work \cite{Price-72}, the author introduced light-cone
variables $u = t - x$ and $v = t + x$, in terms of which
all the wave equations introduced have the same form. We call $V$ the
generic effective potential and $\psi_{\ell}$ the generic field, and the
equations can be written, in terms of the null coordinates, as
\begin{equation}
-4 \frac{\partial^2}{\partial u \partial v} \psi_{\ell} (u,v) =
V (r(u,v)) \psi_{\ell} (u,v) \ .
\label{uv-eq}
\end{equation}
In the characteristic initial value problem, initial data is specified
on the two null surfaces $u = u_{0}$ and $v = v_{0}$. Since the basic
aspects of the field decay are independent of the initial conditions
(this fact is confirmed by our simulations),
we begin with a Gaussian pulse on $u = u_{0}$ and set the field to zero on
$v = v_{0}$, 
\begin{equation}
\psi_{\ell}(u=u_0,v) = \exp\left[-\frac{(v - v_c)^2}{2\sigma^2}\right] \ ,
\end{equation}
\begin{equation}
\psi_{\ell}(u, v=v_0) = \exp\left[-\frac{(v_0 - v_c)^2}{2\sigma^2}\right] .
\end{equation}

Since we do not have analytic solutions to the
time-dependent wave equation with the effective potentials introduced,
one approach is to discretize the equation (\ref{uv-eq}), and
then implement a finite differencing scheme to solve it
numerically. One possible discretization, used for example in
\cite{Wang-01,Brady-97,Brady-99},  is  
\begin{eqnarray}
\lefteqn{\psi_{\ell}(N) = \psi_{\ell}(W) + \psi_{\ell}(E) -
\psi_{\ell}(S) }  \nonumber \\  
& & \mbox{} - \Delta^2 V(S) \frac{ \psi_{\ell}(W) + \psi_{\ell}(E)}{8} +
\mathcal{O}(\Delta^4)   \ , 
\label{d-uv-eq}
\end{eqnarray}
where we have used the definitions for the points: $N = (u + \Delta, v
+ \Delta)$, $W = (u + \Delta, v)$, $E = (u, v + \Delta)$ and $S =
(u,v)$. With the use of expression 
(\ref{d-uv-eq}), the basic algorithm will cover the  region of
interest in the $u-v$ plane, using the value of the field at three points 
in order to calculate  it at a forth one.

After the integration is completed, the values $\psi_{\ell}(u_{max}, v)$
and $\psi_{\ell}(u,v_{max})$ are extracted,  where  $u_{max}$ ($v_{max}$)
is  the  maximum value  of $u$ ($v$)  on  the numerical grid. Taking
sufficiently large $u_{max}$ and $v_{max}$, we have good
approximations for the wave function at the event and cosmological
horizons.

\subsection{Non-characteristic integration}

It is not difficult to set up a numeric algorithm to solve
equation (\ref{wave_equation}) with Cauchy data specified
on a $t$ constant surface. We used $4^{\rm th}$ order in $x$ and
$2^{\rm nd}$ in $t$ scheme (see, for instance,
 \cite{Levander} for an application of this algorithm to seismic
analysis). The second spatial
derivative at a point $(t,x)$, up to $4^{\rm th}$ order, is given by
\begin{eqnarray}
\label{spatial}
\psi_{\ell}''(t,x) & = & \frac{1}{12\Delta x^2}
\left[\psi_{\ell}(t,x+2 \Delta x)  -16\psi_{\ell}(t,x+\Delta x)
\right. \nonumber \\
& & \mbox{} + 30\psi_{\ell}(t,x)- 16\psi_{\ell}(t,x-\Delta x)
\nonumber \\
& &  \left. +\psi_{\ell}(t,x-2\Delta x) \right] \ , 
\end{eqnarray}
while the second time derivative up to $2^{\rm nd}$ order is 
\begin{equation}
\label{time}
\ddot{\psi}_{\ell}(t,x) = \frac{\psi_{\ell}(t+\Delta t,x) -2\psi_{\ell}(t,x) +
\psi_{\ell}(t-\Delta t,x)}{\Delta t^2} \ .
\end{equation}
Given $\psi_{\ell}(t=t_0,x)$ and $\psi_{\ell}=(t=t_0-\Delta t,x)$ 
(or $\dot{\psi}_{\ell}(t=t_0,x)$), we can use (\ref{spatial}) and
(\ref{time}) discretization to solve (\ref{wave_equation}) and
calculate $\psi_{\ell}(t=t_0+\Delta t,x)$. This is the basic
algorithm. At each interaction, one can control the error by using the
invariant integral (the wave energy) associate to (\ref{wave_equation})
\begin{equation}
E = \frac{1}{2}\int \left[ \left( \psi_{\ell}'(t,x)\right)^2
+ \left( \dot{\psi}_{\ell}(t,x)\right)^2 +V(x)
\psi_{\ell}(t,x)^2\right] \, dx \ . 
\end{equation}

We make exhaustive analysis of the asymptotic behavior of the solutions
of (\ref{wave_equation}) with  initial conditions of the
form
\begin{equation}
\psi_{\ell}(0,x) = \exp\left[ -\frac{(x-x_0)^2}{2\sigma_0^2}\right] \ ,
\end{equation}
\begin{equation}
\label{psidot}
\dot{\psi}_{\ell}(0,x) = C\exp\left[ -\frac{(x-x_1)^2}{2\sigma_1^2}\right]
\ . 
\end{equation}
The results do not depend on the details of
initial conditions. They are compatible with the ones obtained by the usual
characteristic  integration, with the only, and significative, exception of
the  $\ell=0$ scalar mode. As we will see, its asymptotic value
depends strongly on the initial velocities $\dot{\psi}_{\ell}(0,x)$. 

\subsection{WKB analysis}

Considering the Laplace transformation of the equation 
 (\ref{wave_equation}), one gets the ordinary
differential equation   
\begin{equation}
\frac{d^2 \psi_{\ell}(x)}{d x^2} - \left[s^2 + V(x) \right]\psi_{\ell}(x) = 0 \ . 
\end{equation}
One finds that there is a discrete set of possible values to $s$
such that the function $\hat{\psi}_{\ell}$, the Laplace transformed
field, satisfies both boundary conditions,
\begin{equation}
\lim_{x \rightarrow -\infty}\hat{\psi}_{\ell} \, e^{sx}=1 \ ,
\end{equation}
\begin{equation}
\lim_{x \rightarrow +\infty}\hat{\psi}_{\ell} \, e^{-sx}=1 \ .
\end{equation}
By making the formal replacement $s=i\omega$, we have the usual quasinormal
mode boundary conditions. The frequencies $\omega$ (or $s$) are
called quasinormal frequencies.

The semi-analytic approach used in this work \cite{Schutz-85,Iyer-87}
is a very efficient algorithm to calculate the quasinormal
frequencies, which have been applied in a variety of situations
\cite{WKB-ap}. With this method, the quasinormal modes are given by
\begin{equation}
\omega_n^2 = \left(V_{0} + P\right) - i\left(n + \frac{1}{2}\right)
            \left(-2V_{0}^{(2)}\right)^{1/2} \left(1 + Q\right)
\label{w-WKB}
\end{equation}
where the quantities $P$ and $Q$ are determined using
\begin{equation}
P = \frac{1}{8} \left[\frac{V_{0}^{(4)}}{V_{0}^{(2)}}\right]
    \left(\frac{1}{4}+\alpha^2\right)
    - \frac{1}{288} \left[\frac{V_{0}^{(3)}}{V_{0}^{(2)}}\right]^{2}
    \left(7 + 60 \alpha^2\right) \ ,
\label{P-WKB}
\end{equation}
\begin{widetext}
\begin{eqnarray}
Q &=& \frac{1}{-2V_{0}^{(2)}} \left\{ 
    \frac{5}{6912} \left[\frac{V_{0}^{(3)}}{V_{0}^{(2)}}\right]^{4}
    \left(77 + 188\alpha^2\right)  
    - \frac{1}{384} \left[\frac{V_{0}^{(3) 2}
    V_{0}^{(4)}}{V_{0}^{(2)}}\right] \left(51 + 100\alpha^2\right)
    \right.    \nonumber  \\ 
   && + \frac{1}{2304} \left[\frac{V_{0}^{(4)}}{V_{0}^{(2)}}\right]^{2}
    \left(67 + 68\alpha^2\right) 
 + \frac{1}{288} \left[\frac{V_{0}^{(3)}
    V_{0}^{(5)}}{V_{0}^{(2) 2}}\right] \left(19 + 28\alpha^2\right) 
   \nonumber  \\
 &&\left. \mbox{} - \frac{1}{288} 
 \left[\frac{V_{0}^{(6)}}{V_{0}^{(2)}}\right] \left(5 + 4\alpha^2\right) 
 \right\} 
\label{Q-WKB}
\end{eqnarray}
\end{widetext}

In the equations (\ref{w-WKB}) to (\ref{Q-WKB}), $\alpha = n + 1/2$ and
the superscript $(i)$ denote differentiation with respect to $x$ of
the potential $V(x)$. The potential and its derivatives are then
calculated in the point $x_{0}$, where $V(x)$ is an extremum.
The integer $n$ labels the modes
\begin{equation}
n = \begin{cases}
    0,1,2,\ldots     & \rm{Re}(\omega)>0  \ ,\\
    -1,-2,-3,\ldots  & \rm{Re}(\omega)<0 \ .
    \end{cases} 
\end{equation}

\section{Near Extreme Limit}

To characterize the near extreme limit of the Schwarzschild-de Sitter
geometry, it is  convenient to define the  dimensionless parameter
$\bar{\delta}$: 
\begin{equation}
\bar{\delta}=\sqrt{1 - 9m^2\Lambda} \ .
\end{equation}
The limit $0<\bar{\delta}\ll1$ is the near extreme limit, where the
horizons  are distinct, but very close. In this regime, analytical
expressions for the frequencies have being calculated
\cite{Cardoso-03,Molina-03}. For the scalar and electromagnetic
fields, the quasinormal frequencies are  
\begin{equation}
\omega_n  = \left[
\frac{\Lambda}{3} - 3m^{2}\Lambda^{2} \right]^{\frac{1}{2}}  \left\{ \left[ 
\ell(\ell + 1) - \frac{1}{4} \right]^{\frac{1}{2}}  - i
\left(n+\frac{1}{2}\right)  \right\} \ .
\label{w-qext-sc}
\end{equation}
For the axial and polar gravitational fields, the frequencies are
given by
\begin{eqnarray}
\omega_n  & = & \left[
\frac{\Lambda}{3} - 3m^{2}\Lambda^{2} \right]^{\frac{1}{2}} \nonumber \\
& & \left\{ \left[ 
(\ell + 2)(\ell - 1) - \frac{1}{4} \right]^{\frac{1}{2}}  - i
\left(n+\frac{1}{2}\right)  \right\} \ .
\label{w-qext-gr}
\end{eqnarray}
They can
be compared with the numerical and semi-analytic methods presented in
the previous section.    
\begin{figure}
\setlength{\unitlength}{1.0mm}
\resizebox{1\linewidth}{!}{\includegraphics*{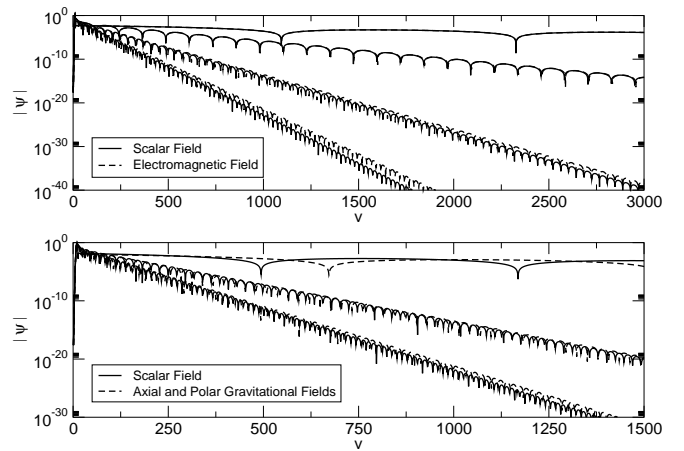}}
\caption{\small \textit{Decay of the scalar and electromagnetic fields with
$\ell=1$, and of the scalar, axial and polar gravitational fields,
with $\ell=2$, with the SdS geometry approaching the near extreme
limit.The parameters for the geometry are $m=1.0$ and
$\bar{\delta}=0.01,0.1,0.3$. }}    
\label{near-ext-l1-2}
\end{figure}
Direct calculation of the wave functions confirms that, in the near
extreme limit, the dynamics is simple, with the late time decay of the
fields being dominated by quasinormal modes. 
All the types of perturbation tend to coincide near the extreme limit.
Besides, as we approach the extreme limit, the oscillation period increases
and the exponential decay rate decreases.
These conclusions, illustrated in figure
\ref{near-ext-l1-2}, for $\ell = 1,2$, 
  are  consistent with the presented in
\cite{Cardoso-03,Molina-03}. 

\begin{figure}[!hc]
\setlength{\unitlength}{1.0mm}
\resizebox{0.8\linewidth}{!}{\includegraphics*{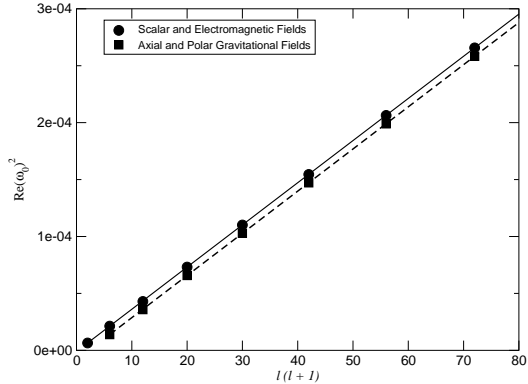}}
\caption{\small \textit{Curve of $Re(\omega_0)^{2}$ with
$\ell(\ell + 1)$, for the scalar, electromagnetic and gravitational
fields, in the near extreme limit. The dots are the numerical
results and the solid lines represent the analytical results.  The
parameters for the geometry are $m=1.0$ and $\bar{\delta}=0.01$.}}  
\label{near-ext}
\end{figure}

By using a non-linear fitting  based in a $\chi^2$ 
analysis, it is possible to estimate the real and
imaginary parts of the $n=0$ quasinormal mode. 
 These results can be compared with the analytical expressions in
the near extreme cases. In the figure \ref{near-ext}, we analyze
the dependence of the frequencies with $\ell$. The accordance is
extremely good.

\section{Intermediary Region in Parameter Space}

\subsection{Scalar Field with $\ell = 0$}

\begin{figure}[!hc]
\setlength{\unitlength}{1.0mm}
\resizebox{0.8\linewidth}{!}{\includegraphics*{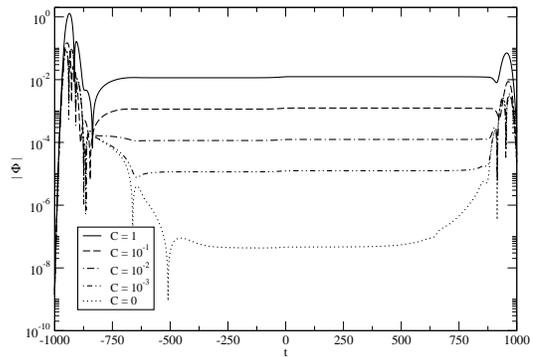}}
\caption{\small \textit{Asymptotic solutions $\Phi(x,t)$ obtained
by non-characteristic numerical integration with $\ell=0$, $\Lambda=10^{-4}$
and $m=1$. The curves correspond to different values of $C$ in
the initial condition (\ref{psidot}).}}  
\label{non-char}
\end{figure}

Only scalar perturbations can have zero total angular momentum.
Solutions of  (\ref{wave_equation}) with $\ell = 0$ leads to  a
constant tail, as already shown in \cite{Brady-97}. This is confirmed
in figure \ref{non-char}. 
The novelty here is the dependence of the asymptotic value
on the $\dot{\psi}_{\ell}(0,x)$ initial condition. 
Figure \ref{non-char} reveals the appearance
of the constant value $\phi_0$ for large $t$, and its dependence on
$\dot{\psi}_{\ell}(0,x)$. Note that $\phi_0$ falls below 
$10^{-7}$ for $\dot{\psi}_{\ell}(0,x)=0$. These results are in
accordance with the analytical predictions of \cite{Brady-99}, which
give
\begin{equation}
\psi \left(\infty,r\right) = \frac{r}{r_c^2}\int_0^{r_c}
\dot{\psi}(0,s)s\frac{ds}{h'(s)}.
\end{equation}

\subsection{Fields with $\ell > 0$ }

We can have scalar and vector fields with angular momentum
$\ell=1$, and with $\ell > 1$, it is possible to introduce also
gravitational fields.  Their behavior is described
in general by three phases. The first
corresponds to the quasinormal modes generated from the presence of the
black hole itself. A little later there is a region of power-law decay,
which continues indefinitely in an asymptotic flat space. In the presence of a
positive cosmological constant, however, an exponential decay takes
over in the latest period.

Some qualitatively different effects show up when we turn away from the
near extreme limit. For a small cosmological constant the asymptotic
behavior is dominated by an exponential decaying mode rather than by a
quasinormal mode. Such exponential decays will
characterize the de Sitter space in general. We will later comment about
the pure de Sitter limit, which is exactly solvable, though we do not have
definite answers about the respective behavior. The exponential decay
also characterizes the electromagnetic as well as the gravitational
perturbations.

\begin{figure}
\setlength{\unitlength}{1.0mm}
\resizebox{1.0\linewidth}{!}{\includegraphics*{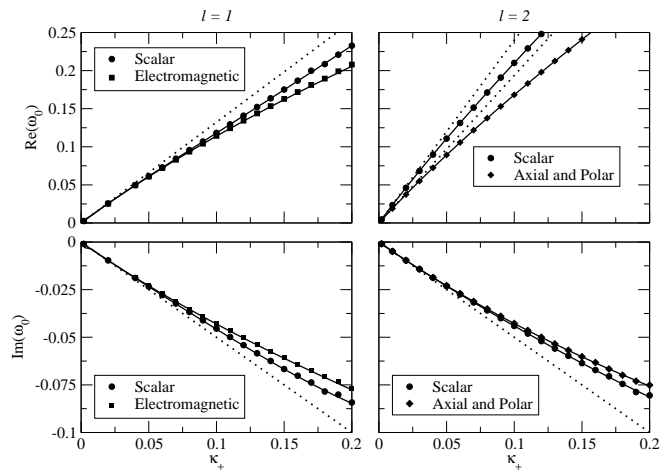}}
\caption{\small \textit{Graphs of the real and imaginary part of the
fundamental frequencies $(n=0)$ with $\kappa_{+}$. The dotted
lines are the near extreme results, the dots are the numerical
results and the continuous curves are the semi-analytic results. In
the graphs, $m=1.0$.}}  
\label{w-kappap}
\end{figure}

As the separation of the horizons increases, the quasinormal
frequencies deviate from the predicted by expressions
(\ref{w-qext-sc}) and (\ref{w-qext-gr}). In figure \ref{w-kappap},
this is illustrated for $\ell=1,2$. 

It is interesting to compare the values obtained for the fundamental
modes using the numerical and semi-analytic methods. We find that the agreement
between then is good, for the hole range of $\Lambda$. The difference
is better for the first values of $\ell$. 
This is expected, since the numerical calculations are better in this
region.  In the table \ref{SC-num-WKB}, we illustrate
these observations with a few values of $\Lambda$. It is important to
mention that quasinormal frequencies for the SdS black hole were
already calculated in a recent paper \cite{Zhidenko}, applying a
variation of the WKB method used here \cite{Konoplya-03}.  There are earlier
papers calculating quasinormal modes in this geometry, for example
\cite{dsqnm}.    

\begin{table}
\begin{tabular}{|c|c||c|c||c|c|}
\cline{3-6} 
\multicolumn{2}{c||}{}         &
\multicolumn{2}{c||}{Numerical}&
\multicolumn{2}{c|}{Semi-analytical}\\
\hline 
$\ell$ & $\Lambda$ &Re($\omega$)& -Im($\omega$)&Re($\omega$)& -Im($\omega$)\\
\hline \hline 
1  & 1.000E-5 & 2.930E-01 & 9.753E-01 & 2.911E-01 & 9.780E-02 \\
\hline 
   & 1.000E-4 & 2.928E-01 & 9.764E-02 & 2.910E-01 & 9.797E-02 \\
\hline 
   & 1.000E-3 & 2.914E-01 & 9.726E-02 & 2.896E-01 & 9.771E-02 \\
\hline 
   & 1.000E-2 & 2.770E-01 & 9.455E-02 & 2.753E-01 & 9.490E-02 \\
\hline 
   & 1.000E-1 & 8.159E-02 & 3.123E-02 & 8.144E-02 & 3.137E-02 \\
\hline \hline 
2  & 1.000E-5 & 4.840E-01 & 9.653E-02 & 4.832E-01 & 9.680E-02 \\
\hline 
   & 1.000E-4 & 4.833E-01 & 8.948E-02 & 4.830E-01 & 9.677E-02 \\
\hline 
   & 1.000E-3 & 4.816E-01 & 8.998E-02 & 4.809E-01 & 9.643E-02 \\ 
\hline 
   & 1.000E-2 & 4.598E-01 & 8.880E-02 & 4.592E-01 & 9.290E-02 \\
\hline 
   & 1.000E-1 & 1.466E-01 & 3.068E-02 & 1.466E-01 & 3.070E-02 \\
\hline \hline  
3  & 1.000E-5 & 6.769E-01 & 8.662E-02 & 6.752E-01 & 9.651E-02 \\
\hline 
   & 1.000E-4 & 6.754E-01 & 8.654E-02 & 6.749E-01 & 9.647E-02 \\
\hline 
   & 1.000E-3 & 6.732E-01 & 8.660E-02 & 6.720E-01 & 9.611E-02 \\
\hline 
   & 1.000E-2 & 6.437E-02 & 9.200E-02 & 6.428E-02 & 9.235E-02 \\
\hline 
   & 1.000E-1 & 2.091E-02 & 3.054E-02 & 2.091E-02 & 3.056E-02 \\
\hline
\end{tabular}
\caption{\small \textit{Fundamental frequency ($n=0$) of the scalar
field, obtained using the numerical and semi-analytical methods. In
this table, $m=1.0$.}}
\label{SC-num-WKB}
\end{table}

\begin{table}
\begin{tabular}{|c|c||c|c||c|c|}
\cline{3-6} 
\multicolumn{2}{c||}{}         &
\multicolumn{2}{c||}{Numerical}&
\multicolumn{2}{c|}{Semi-analytical}\\
\hline 
$\ell$ & $\Lambda$ &Re($\omega$)& -Im($\omega$)&Re($\omega$)& -Im($\omega$)\\
\hline \hline 
1  & 1.000E-5 & 2.481E-01 & 9.226E-02 & 2.459E-01 & 9.310E-02 \\
\hline 
   & 1.000E-4 & 2.481E-01 & 9.223E-02 & 2.457E-01 & 9.307E-02 \\
\hline 
   & 1.000E-3 & 2.475E-01 & 9.176E-02 & 2.448E-01 & 9.270E-02 \\
\hline 
   & 1.000E-2 & 2.374E-01 & 8.839E-02 & 2.352E-01 & 8.896E-02 \\
\hline 
   & 1.000E-1 & 8.035E-02 & 3.027E-02 & 8.023E-02 & 3.033E-02 \\
\hline \hline 
2  & 1.000E-5 & 4.577E-01 & 8.985E-02 & 4.571E-01 & 9.506E-02 \\
\hline 
   & 1.000E-4 & 4.575E-01 & 8.991E-02 & 4.569E-01 & 9.502E-02 \\
\hline 
   & 1.000E-3 & 4.559E-01 & 9.439E-02 & 4.551E-01 & 9.464E-02 \\ 
\hline 
   & 1.000E-2 & 4.371E-01 & 8.941E-02 & 4.364E-01 & 9.074E-02 \\
\hline 
   & 1.000E-1 & 1.458E-01 & 3.037E-02 & 1.458E-01 & 3.038E-02 \\
\hline \hline  
3  & 1.000E-5 & 6.578E-01 & 8.365E-02 & 6.567E-01 & 9.563E-02 \\
\hline 
   & 1.000E-4 & 6.576E-01 & 8.349E-02 & 6.564E-01 & 9.559E-02 \\
\hline 
   & 1.000E-3 & 6.547E-01 & 8.399E-02 & 6.538E-01 & 9.520E-02 \\
\hline 
   & 1.000E-2 & 6.276E-02 & 8.852E-02 & 6.267E-01 & 9.125E-02 \\
\hline   
   & 1.000E-1 & 2.085E-02 & 3.039E-03 & 2.085E-01 & 3.040E-02 \\
\hline
\end{tabular}
\caption{\small \textit{Fundamental frequency ($n=0$) of the
electromagnetic field, obtained using the numerical and
semi-analytical methods. In this table, $m=1.0$.}}  
\label{EL-num-WKB}
\end{table}

\begin{table}
\begin{tabular}{|c|c||c|c||c|c|}
\cline{3-6} 
\multicolumn{2}{c||}{}         &
\multicolumn{2}{c||}{Numerical}&
\multicolumn{2}{c|}{Semi-analytical}\\
\hline 
$\ell$ & $\Lambda$ &Re($\omega$)& -Im($\omega$)&Re($\omega$)& -Im($\omega$)\\
\hline \hline 
2  & 1.000E-5 & 3.738E-01 & 8.883E-02 & 3.731E-01 & 8.921E-02 \\
\hline
   & 1.000E-4 & 3.737E-01 & 8.880E-02 & 3.730E-01 & 8.918E-02 \\
\hline 
   & 1.000E-3 & 3.721E-01 & 8.850E-02 & 3.715E-01 & 8.888E-02 \\ 
\hline 
   & 1.000E-2 & 3.566E-01 & 8.538E-02 & 3.560E-01 & 8.572E-02 \\
\hline 
   & 1.000E-1 & 1.179E-01 & 3.020E-02 & 1.179E-01 & 3.023E-02 \\
\hline \hline 
3  & 1.000E-5 & 5.999E-01 & 8.677E-02 & 5.992E-01 & 9.272E-02 \\
\hline 
   & 1.000E-4 & 5.996E-01 & 8.676E-02 & 5.990E-01 & 9.269E-02 \\
\hline 
   & 1.000E-3 & 5.972E-01 & 8.971E-02 & 5.966E-01 & 9.234E-02 \\
\hline 
   & 1.000E-2 & 5.725E-01 & 8.695E-02 & 5.718E-01 & 8.874E-02 \\
\hline 
   & 1.000E-1 & 1.900E-01 & 3.030E-02 & 1.900E-02 & 3.032E-02 \\
\hline \hline  
4  & 1.000E-5 & 8.106E-01 & 8.810E-02 & 8.091E-01 & 9.417E-02 \\
\hline 
   & 1.000E-4 & 8.102E-01 & 8.781E-02 & 8.087E-01 & 9.413E-02 \\
\hline 
   & 1.000E-3 & 8.070E-01 & 8.799E-02 & 8.055E-01 & 9.376E-02 \\
\hline 
   & 1.000E-2 & 7.733E-02 & 8.714E-02 & 7.720E-01 & 9.000E-02 \\
\hline 
   & 1.000E-1 & 2.564E-02 & 3.034E-03 & 2.563E-01 & 3.036E-02 \\
\hline
\end{tabular}
\caption{\small \textit{Fundamental frequency ($n=0$) of the axial and polar
gravitational fields, obtained using the numerical and semi-analytical
methods. In this table, $m=1.0$.} }
\label{AX-num-WKB}
\end{table}

\begin{figure}
\setlength{\unitlength}{1.0mm}
\resizebox{1\linewidth}{!}{\includegraphics*{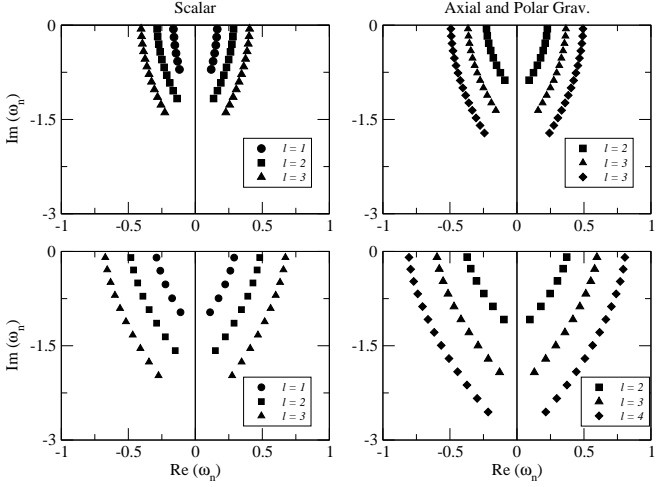}}
\caption{\small \textit{Quasinormal modes of the scalar, axial and polar
gravitational fields, for higher modes. The parameter for the curves
are $m=1$ and $\Lambda=10^{-3}$.}}  
\label{multi-n}
\end{figure}

The first higher $n$ modes cannot be obtained from the numerical
solution, but can be calculated by the semi-analytical method. As the
cosmological constant decreases, the real and imaginary parts of the
frequencies increase, up to the limit where  the geometry is
asymptotically flat. The behavior of the modes is illustrated in
figure \ref{multi-n}. The behavior of the electromagnetic field is
similar.   

\begin{figure}
\setlength{\unitlength}{1.0mm}
\resizebox{1\linewidth}{!}{\includegraphics*{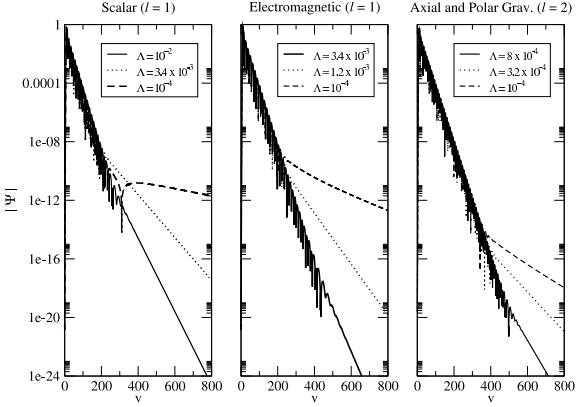}}
\caption{\small \textit{Exponential tail for the scalar and
electromagnetic fields with $\ell=1$, and for the axial and polar
gravitational field with $\ell=2$. In the graphs, $m=1.0$.  }}  
\label{exp-dec}
\end{figure}

A $\chi^2$ analysis of the data presented in 
figure \ref{exp-dec} shows that the massless scalar, electromagnetic
and gravitational perturbations in SdS geometry behave as
\begin{eqnarray}
\psi_{\ell}^{sc}\sim e^{-k_{exp}^{sc}t} & {\rm with} &
t\rightarrow\infty
\end{eqnarray}
\begin{eqnarray}
\psi_{\ell}^{el}\sim e^{-k_{exp}^{el}t} & {\rm with} &
t\rightarrow\infty
\end{eqnarray}
\begin{eqnarray}
\psi_{\ell}^{ax}\sim e^{-k_{exp}^{ax}t} & {\rm with} &
t\rightarrow\infty
\end{eqnarray}
\begin{eqnarray}
\psi_{\ell}^{po}\sim e^{-k_{exp}^{po}t} & {\rm with} &
t\rightarrow\infty
\end{eqnarray}
for $t$ sufficiently large. At the event and the cosmological horizons
$t$ is substituted, respectively by  $v$ and $u$.

\begin{figure}
\setlength{\unitlength}{1.0mm}
\resizebox{1\linewidth}{!}{\includegraphics*{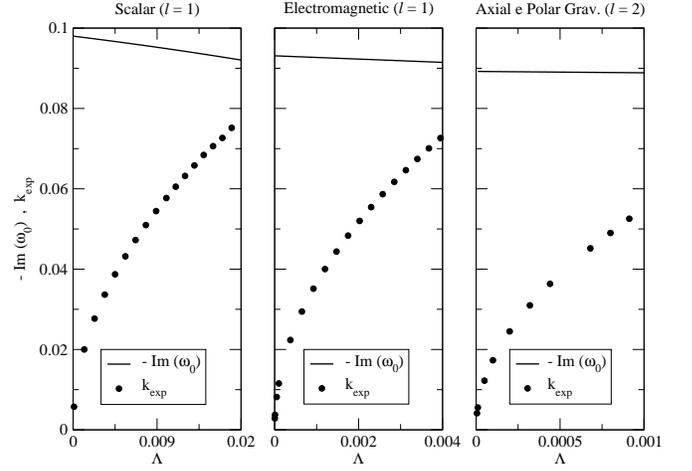}}
\caption{\small \textit{Approaching of the constants $-\textrm{Im}
(\omega_0)$ and $k_{exp}$, for the scalar, electromagnetic and
gravitational fields, in the SdS geometry. Above a certain critical
value of $\Lambda$ (roughly $1.7 \times 10^{-2}$, $4.0 \times
10^{-3}$ and $1.2 \times 10^{-3}$ respectively, for the parameters taken in
graphs), a tail it is not observed. For all curves, the mass is set to
$m=1.0$. }} 
\label{kexp-wI}
\end{figure}

The numerical simulations developed in the present work reveal an
interesting transition between oscillatory modes and exponentially decaying
modes. As shown in figure \ref{kexp-wI}, as the cosmological
constant decreases, the absolute value of  $-\textrm{Im}
(\omega)$ decreases.

Above a certain critical value of $\Lambda$ we do not observe the
exponential tail, since the coefficient  $k_{exp}$ is larger than
$-\textrm{Im}(\omega_0)$  thus the decaying quasinormal mode
dominates. But for $\Lambda$ smaller than this critical value,
$-\textrm{Im}(\omega_0)$ turns out to be larger than $k_{exp}$ and the
exponential tail dominates. Certainly, for a small enough cosmological
constant the exponential tail dominates in the various cases
considered here.

\begin{figure}
\setlength{\unitlength}{1.0mm}    
\resizebox{1\linewidth}{!}{\includegraphics*{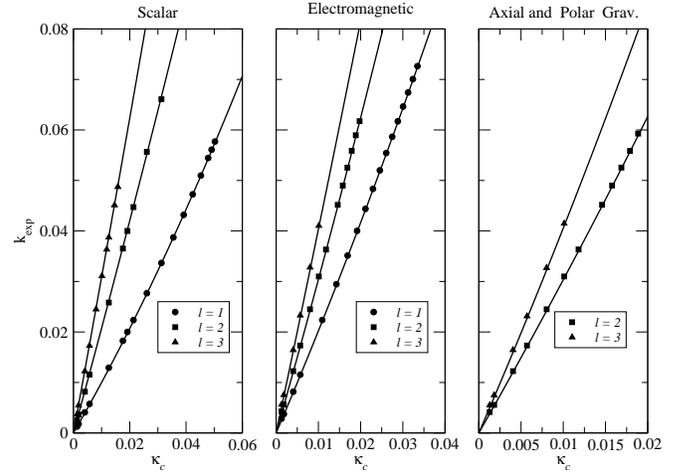}}
\caption{\small \textit{Dependence of  $k_{exp}$ with
$\kappa_{c}$ and $\ell$, in the SdS geometry.  The symbols indicate
the numerical values, and the solid lines are the appropriate fittings. 
For the left graph: 
$k_{exp}^{sc}=1.077\times 10^{-4} + 0.984\kappa_{c} + 3.545\kappa_{c}^2$,  
$k_{exp}^{sc}=-1.863\times 10^{-4} + 2.010\kappa_{c} + 2.608\kappa_{c}^2$ and 
$k_{exp}^{sc}=-1.959\times 10^{-4} + 3.028\kappa_{c} + 2.978\kappa_{c}^2$. 
For the center graph: 
$k_{exp}^{el} = 2.082 \times 10^{-4} + 1.988\kappa_{c} + 6.141\kappa_{c}^2$,  
$k_{exp}^{el} = 2.712 \times 10^{-4} + 2.974\kappa_{c} + 8.284\kappa_{c}^2$ and 
$k_{exp}^{el} = 3.737 \times 10^{-4} + 3.984\kappa_{c} + 4.517\kappa_{c}^2$. 
For the right graph: 
$k_{exp}^{ax} = 2.616 \times 10^{-4}+ 2.974\kappa_{c} +
9.895\kappa_{c}^2$ and
$k_{exp}^{ax} = 4.484\times 10^{-4} + 3.896\kappa_{c} +
18.92\kappa_{c}^2$. In the graphs, $m=1.0$.}}   
\label{kexp-kappac}
\end{figure}

Another aspect worth mentioning in the intermediate region is the
dependence of the parameters  $k_{exp}^{sc}$, $k_{exp}^{el}$,
$k_{exp}^{ax}$ and $k_{exp}^{po}$ with  $\ell$ and $\kappa_{c}$. The
results suggest that the $k_{exp}$ are at least second differentiable
functions of $\kappa_{c}$. Therefore, close to $\kappa_{c}=0$, we
approximate 
\begin{equation}
k_{exp}^{sc}(\kappa_{c})\approx\ell \left(\kappa_{c}+c^{sc}
\kappa_{c}^{2}\right) \ ,
\end{equation}
\begin{equation}
k_{exp}^{el}(\kappa_{c}) \approx k_{exp}^{ax}(\kappa_{c}) \approx 
k_{exp}^{po}(\kappa_{c}) \approx(\ell+1)
\left(\kappa_{c}+c^{e-g}\kappa_{c}^{2}\right) \ . 
\end{equation}
Previous results are illustrated in figure \ref{kexp-kappac}. 

\section{Approaching the Asymptotically Flat Geometry}

Scalar fields in the SdS geometry near the asymptotically flat limit
wore studied in \cite{Brady-97,Brady-99}. In this case there is
a clear separation between the event and the cosmological horizons,
such that 
\begin{equation}
\delta=\frac{r_{c}-r_{+}}{r_{+}} \gtrsim 50  \ .
\end{equation}

\begin{figure}
\setlength{\unitlength}{1.0mm}    
\resizebox{1\linewidth}{!}{\includegraphics*{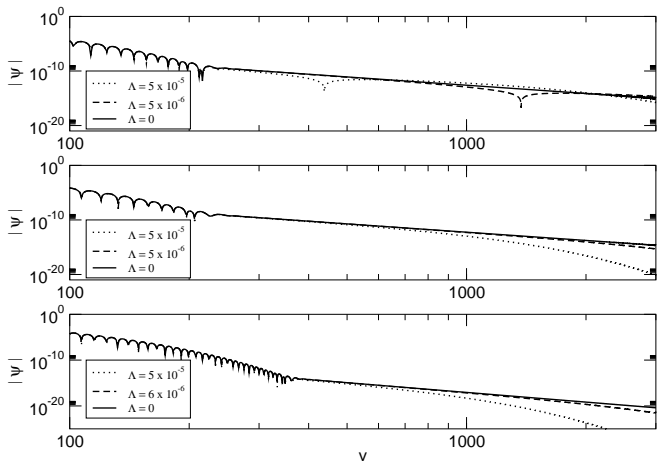}}
\caption{\small \textit{Approaching the asymptotically flat
geometry. Straight lines in the log-log graphs indicate power-law
decay. In the graphs, $m=1.0$. }}   
\label{appr-S}
\end{figure}

A new qualitative change occurs in this regime, namely a decaying
phase with a power-law behavior. Such a phase occurs between the
quasinormal mode decay and the exponential decay phases. The field
cannot be simply described by a superposition of the various modes,
which would imply a domination of the power-law phase. This is
illustrated in figure \ref{appr-S}.

\section{Conclusions}

We have identified three regimes, according to the value of $\Lambda$ 
 for the decay of the scalar,
electromagnetic and gravitational perturbations. 
Near the extreme limit (high $\Lambda$), 
we have analytic expressions for the
effective potentials and to the quasinormal frequencies. The decay is
entirely dominated by the quasinormal modes, 
that is, oscillatory decaying characterized by a
 non vanishing real part of the quasinormal frequency.

In an intermediary parameter region (lower $\Lambda$), the wave
functions have an important qualitative change, with the appearance
of an exponential tail. This tail dominates the decay for large
time. Near the asymptotically flat limit ($\Lambda \ll 1$), we see an
intermediary phase between the quasinormal modes and the exponential
tail --- a region of power-law decay. When $\Lambda=0$, this region entirely
dominates the late time behavior.  

Finally, for scalar fields with $\ell = 0$ a constant 
decay mode appears, and its value 
depends on the $\dot{\psi}_{\ell}(0,x)$ initial condition. Figure
\ref{non-char} reveals the appearance of the constant value $\phi_0$
for large $t$, and its dependence on $\dot{\psi}_{\ell}(0,x)$. The
value of $\phi_0$ falls below  $10^{-7}$ for
$\dot{\psi}_{\ell}(0,x)=0$.  

An immediate extension of the work developed in this paper is the in
treatment of field propagation in the exterior of charged and
asymptotically de Sitter  black holes. In this case, the
electromagnetic and gravitational perturbations are necessarily
coupled. It would be interesting to see if the general picture
presented here is still valid in this more general context.


\begin{acknowledgments}
This work was supported by Funda\c{c}\~{a}o de Amparo
\`{a} Pesquisa do Estado de S\~{a}o Paulo (FAPESP) and by Conselho
Nacional de Desenvolvimento Cient\'{\i}fico e Tecnol\'{o}gico (CNPq),
Brazil. 
\end{acknowledgments}



\end{document}